\documentclass[a4paper,12pt]{article}
\usepackage{graphicx}
\usepackage{epsfig}
\usepackage[american]{babel}

\begin{document}
\title{Scaling properties in dynamics of non-analytic complex maps near the
accumulation point of the period-tripling cascade}

\author{O.B.~Isaeva, S.P.~Kuznetsov}
\date{}
\maketitle\begin{center} \emph{Institute of Radio-Engineering and
Electronics of RAS, \\ Zelenaya 38, Saratov, 410019,
Russia}\end{center}

\maketitle\begin{center} \emph{Saratov State University, \\
Astrakhanskaya 83, Saratov, 410026, Russia}\end{center}

\maketitle
\begin{abstract}
The accumulation point of the period-tripling bifurcation cascade
in complex quadratic map was discovered by Golberg, Sinai, and
Khanin (Russ.Math.Surv. 38:1, 1983, 187), and independently by
Cvitanovi\'c and Myrheim (Phys.Lett. A94:8, 1983, 329). As we
argue, in the extended parameter space of smooth maps, not
necessary satisfying the Cauchy -- Riemann equations, the scaling
properties associated with the period-tripling are governed by two
relevant universal complex constants. The first one is
$\delta_{1}\cong 4.6002-8.9812i$ (in accordance with the mentioned
works), while the other one, responsible for the violation of the
analyticity, is found to be $\delta_{2}\cong 2.5872+1.8067i$. It
means that in the extended parameter space the critical behaviour
associated with the period-tripling cascade is a phenomenon of
codimension four. Scaling properties of the parameter space are
illustrated by diagrams in special local coordinates. We emphasize
a necessity for a nonlinear parameter change to observe the
parameter space scaling.

\end{abstract}
\section{Introduction}
One of the most popular illustrations in the nonlinear science is
a picture of the Mandelbrot set \cite{Peitgen,Devaney} (Fig. 1).
On the complex parameter plane it is defined as the set of points
$\lambda$, at which the iterations of the complex quadratic map
\begin{equation}\label{logistic}
 z'=\lambda-z^{2}, \quad \lambda,z \in \mathrm{C}
\end{equation}
launched from the origin $z=0$ never diverge. (Here the prime
marks the dynamical variable relating to the next iteration, i.e.
to the next moment of discrete time.)

The Mandelbrot set has subtle and complicated structure, which is
a subject of extensive researches. Objects analogous to the
Mandelbrot set are present in parameter planes of other complex
analytic iterative maps too \cite{Devaney,Briggs}.

The Mandelbrot set is regarded as a classic example of fractal;
this suggests a sort of self-similarity, or scaling.\footnote{In
fact, as shown by Milnor \cite{Milnor}, the subtle structure
("hairiness") of the Mandelbrot set does not reproduce itself on
deep levels of resolution in small-scales. Speaking more
accurately, one should relate the property of self-similarity
rather to a definite subset called the Mandelbrot cactus
\cite{Cvitanovic1}. It consists of a domain of existence of a
stable fixed point and of domains associated with attractive
periodic orbits, which originate from the fixed point via all
possible bifurcation sequences (see the grey colored part of the
picture in Fig. 1).} One particular manifestation of scaling is
the Feigenbaum period-doubling cascade. The leaves of the
Mandelbrot cactus placed along the real axis correspond to periods
$2$, $4$, $8$,...$2^{k}$..., and, in asymptotics of large $k$,
they perfectly reproduce each other under subsequent magnification
by a real universal factor $\delta_{F}\cong 4.6692$. The limit
point of the period-doubling accumulation is disposed on the real
axis at $\lambda_{F}=1.401155...$, and it is called the Feigenbaum
critical point. The scaling properties follow from the
renormalization group (RG) analysis developed by Feigenbaum
\cite{Feigenbaum1,Feigenbaum2}. Behaviour of the solution of the
RG equation in complex domain was discussed e.g. by Nauenberg
\cite{Nauenberg} and by Wells and Overill \cite{Wells}.

One can find many other critical points in the parameter plane of
the map (\ref{logistic}), at which the Mandelbrot cactus displays
other scaling regularities. In particular, we can select a path on
the complex plane across the sequence of leaves corresponding to
periods $3$, $9$, $27$..., $3^{k}$..., and arrive at the
period-tripling accumulation point
\begin{equation}\label{pta}
\lambda=0.0236411685377+0.7836606508052i.
\end{equation}
As this point was first discovered by Golberg, Sinai, and Khanin
\cite{Golberg}, we call it the {\it GSK critical point}.
Independently, it was found and studied also by Cvitanovi\'c and
Myrheim \cite{Cvitanovic1,Cvitanovic2}. As in the case of
period-doubling, the RG analysis has been applied, and
self-similarity of the sequence of leaves of the Mandelbrot cactus
has been revealed. According to
Refs.~\cite{Golberg,Cvitanovic2,Cvitanovic1}, the scaling factor
in the case of period-tripling is complex, $\delta_{1}\cong
4.6002-8.9812i$.\footnote{Beside (\ref{pta}), a complex conjugate
period-tripling accumulation point exists. It is placed at
$\lambda=\lambda_{c}^{*}$, and scaling in its neighborhood is
governed by the conjugate constant $\delta_{1}^{*}$.}

Sometimes it may be preferable to deal not with the
one-dimensional complex analytic maps, but with the equivalent
class of two-dimensional real maps satisfying the Cauchy --
Riemann equations.

It would be interesting to discuss a possibility of observation of
phenomena associated with complex analytic dynamics and the
Mandelbrot set in physical systems. Recently this question was
posed by Beck \cite{Beck}. He has considered motion of a particle
in a double-well potential in a time-depended magnetic field. As
he has stated, under certain assumptions dynamics of the particle
may be described by the complex quadratic map or by other complex
analytic maps. It should be emphasized, however, that discussing
physical applications for any special type of dynamics we should
take into account a robustness of a phenomenon under study. In
this context it is important to know either phenomena associated
with the complex analytic dynamics survive or not in slightly
modified maps, which can violate the Cauchy -- Riemann conditions.

In the original work \cite{Golberg} the authors claimed that the
infinite period-tripling cascade could occur typically in
two-parameter families of real two-dimensional maps (see also
\cite{Gunaratne}). In contrast, from studies of Peinke et al.
\cite{Peinke}, Klein \cite{Klein}, and from qualitative analysis
of Cvitanovi\'c and Myrheim \cite{Cvitanovic1} it follows that in
non-analytic maps the parameter space arrangement is changed
drastically, and, apparently, there is no opportunity of surviving
for the universality intrinsic to the period-tripling. Recently
Peckham and Montaldi have presented extensive bifurcation analysis
of a complex quadratic map with a non-analytic term
\cite{Peckham1,Peckham2}. However, the question what happens with
the bifurcation cascades of period-tripling remains not
clear.\footnote{The same assertion relates to other multi-tupling
bifurcation cascades, although we do not touch them in the present
paper.} Sure enough, this matter deserves accurate quantitative
analysis in terms of the RG approach, and this is the main goal of
the present article. Our final conclusion is that in real
two-dimensional maps the critical behaviour associated with the
infinite period-tripling cascade will occur as a phenomenon of
codimension 4, i.e. it may be found at some special points of a
four-dimensional parameter space. In other words, the codimension
is so high that it seems not easy to observe the phenomenon in
physical systems, at least in absence of some special symmetries.

In section 2 we recall contents of the RG analysis developed in
\cite{Golberg,Cvitanovic2,Cvitanovic1}. Then, it is extended to
study the fixed point solution of the RG equation in respect to a
class of perturbations, which can violate the Cauchy -- Riemann
conditions. We find that among the non-analytic eigenmodes of the
linearized RG transformation one is of relevance. In section 3 we
formulate the model map appropriate for a study of the extended
parameter space. Then we explain a necessity of nonlinear variable
change to define local coordinates for observation of scaling in
the parameter space, and construct a procedure to find this
coordinate change numerically. In section 4 we discuss some
details of the parameter space arrangement near the GSK critical
point and present graphical illustrations of the intrinsic scaling
properties.
\section{Renormalization group analysis}
To start, let us follow
Refs.~\cite{Golberg,Cvitanovic2,Cvitanovic1} and assume that
discrete time evolution of the complex variable $z$ is governed by
the complex quadratic map (\ref{logistic}). Then, for three
iterations we have
\begin{equation}\label{f3}
z'=f(f(f(z)))=\lambda-(\lambda-(\lambda-z^{2})^{2})^{2}.
\end{equation}
Next, we rescale the variable $z$ by a factor $\alpha_{0}$ and
choose this factor to normalize the new mapping to unity at the
origin. Then we have
\begin{equation}\label{z3}
z'=f_{1}(z),
\end{equation}
where
\begin{equation}\label{f1}
f_{1}(z)=\alpha_{0}f(f(f(z/\alpha_{0}))),\quad \mbox{and} \quad
\alpha_{0}=1/f(f(f(0))).
\end{equation}
This is just one step of the RG transformation.

Now we can apply the same procedure to the function $f_{1}(z)$ and
obtain the next function $f_{2}(z)$, and so on. By multiple
repetition of the transformation we arrive at the recurrent
functional equation
\begin{equation}\label{RG}
 f_{k+1}(z)=\alpha_{k}f_{k}(f_{k}(f_{k}(z/\alpha_{k}))),
\qquad
 \alpha_{k}=1/f_{k}(f_{k}(f_{k}(0))).
\end{equation}
Here $f_{k}(z)$ represents the evolution operator for $3^{k}$
iterations of the original map in terms of the renormalized
dynamical variable. Note that the rescaling constants
$\alpha_{k}$, which appear at the subsequent steps of the
procedure, are complex.

According to Refs.~\cite{Golberg,Cvitanovic2,Cvitanovic1}, at the
GSK critical point the sequence $f_{k}(z)$ converges to the fixed
point function $g(z)=\lim\limits_{k\to\infty} f_{k}(z)$, which
obeys the equation
\begin{equation}\label{RG1}
g(z)=\alpha g(g(g(z/\alpha))), \quad
\alpha=\lim\limits_{k\to\infty} \alpha_{k}=1/g(g(g(0))).
\end{equation}
It is a generalization of the Feigenbaum-Cvitanovi\'c equation
\cite{Feigenbaum1,Feigenbaum2} for the case of period-tripling.
This functional equation may be solved numerically by means of
approximating the function $g(z)$ via a finite polynomial
expansion and computing the coefficients with a use of
multi-dimensional Newton method \cite{Golberg,Cvitanovic2}. We
have reproduced these calculations, and present the results in
Table I. Note that only terms of even power are nonzero.

As follows from the computations, the scaling constant is
\begin{equation}\label{alfa}
\alpha=-2.09691989+2.35827964i
\end{equation}
in good agreement with Refs. \cite{Golberg,Cvitanovic2,
Cvitanovic1}.

As the rescaling constant is known, it is convenient to redefine
the RG transformation. Now we will assume that the rescaling
factor is the same at all subsequent steps of the procedure and
equals to the universal value (\ref{alfa}). We just substitute
$\alpha_{k}=\alpha$ into Eq.(\ref{RG}) and obtain
\begin{equation}\label{fa}
f_{k+1}(z)=\alpha f_{k}(f_{k}(f_{k}(z/ \alpha))).
\end{equation}
Obviously, this version of the RG transformation possesses the
same fixed point $g(z)$, although the normalization $g(0)=1$
should be regarded now as an arbitrarily accepted additional
condition.

Let us consider the map $z'=g(z)$ corresponding to the fixed-point
solution of the RG equation. It may be checked numerically that
this map has an unstable fixed point $z=z_{*}\cong
0.691473-0.302692i$, and derivative at this point is
\begin{equation}\label{mu}
\mu_{c}=g'(z_{*}) =-0.47653180-1.05480868i.
\end{equation}
Then, as follows from Eq.~(\ref{RG1}), $z_{*}/\alpha$ is a point
belonging to an orbit of period $3$. The Floquet eigenvalue (or
multiplier) for this cycle will be the same complex number
$\mu_{c}$. By induction, it is easy to prove that there is an
infinite countable set of the period-$3^{k}$ cycles
($z_{*}/\alpha^k$ belongs to the period-$3^k$ orbit), with the
same value of the multiplier.

As the mapping $z'=g(z)$ represents the asymptotic form of the
evolution operators for the original map (\ref{logistic}), it must
have an infinite set of unstable orbits of periods $3^{k}$ at the
GSK critical point. The same is true for any other map relating to
the same universality class. The asymptotic value of the
multipliers for the set of cycles is the universal constant
$\mu_{c}$.

Now let us assume that we have a real smooth two-dimensional map
\begin{equation}\label{XY}
x'=\varphi (x,y), \quad y'=\psi (x,y),
\end{equation}
which depends on some parameters. In terms of the complex variable
this map may be expressed, in general, via a function of two
arguments $z=x+iy$ and $z^{*}=x-iy$: $z'=F(z,z^{*})$, where
$F(x+iy, x-iy)=\varphi (x,y)+i \psi (x,y)$. Next, we suppose that
at some special values of parameters the map becomes complex
analytic, i.e. the Cauchy -- Riemann equations hold:
$\varphi_{x}=\psi_{y}$, $\varphi_{y}=-\psi_{x}$, and the dynamics
is just of the kind intrinsic to the GSK critical point. It means
that for a number of iterations $3^{k}$ ($k$ large) the evolution
operator is represented in appropriate normalization by the
universal function $g(z)$. Now, if we slightly change the values
of parameters and depart from the original point in the parameter
space, some perturbation to the fixed point of the RG
transformation will appear. In general, the perturbation may not
satisfy the Cauchy -- Riemann conditions, so the evolution
operator for $3^k$ steps of iterations must be written as
\begin{equation}\label{fgh}
F_{k}(z)=g(z)+\epsilon h_{k}(z,z^{*}).
\end{equation}
Here $\epsilon \ll 1$, and $h_{k}$ is a smooth function of two
arguments. By three-fold iteration of the map (\ref{fgh}) with
rescaling of $z$ by factor $\alpha$ we arrive (in the first order
in $\epsilon$) to the same form of the operator, but with the
modified perturbation term:
\begin{equation}\label{iterh}
\begin{array}{c}
h_{k+1}(z,z^{*})=\alpha[g'(g(g(z/\alpha)))g'(g(z/\alpha))h_{k}(z/\alpha,(z/\alpha)^{*})+
\\  +g'(g(g(z/\alpha)))h_{k}(g(z/\alpha),(g(z/\alpha))^{*})+
h_{k}(g(g(z/\alpha)),(g(g(z/\alpha)))^{*})],
\end{array}
\end{equation}
where $g'$ means the derivative of the function $g$. Obviously,
this relation has a structure
\begin{equation}\label{oper}
h_{k+1}=\hat m h_{k}
\end{equation}
where $\hat m$ is a linear operator defined by the right-hand part
of (\ref{iterh}).

The behaviour of $h_{k}$ under subsequent iterations of the linear
operator is linked with the spectrum of eigenvalues. Indeed, let
we have an eigenfunction $h(z,z^{*})$ associated with an
eigenvalue $\nu$:
\begin{equation}\label{iterhnu}
\begin{array}{c}
\nu
h(z,z^{*})=\alpha[g'(g(g(z/\alpha)))g'(g(z/\alpha))h(z/\alpha,(z/\alpha)^{*})+
\\  +g'(g(g(z/\alpha)))h(g(z/\alpha),(g(z/\alpha))^{*})+
h(g(g(z/\alpha)),(g(g(z/\alpha)))^{*})].
\end{array}
\end{equation}
Then, the functional sequence $h_{k}$ may contain a component,
which behaves as $\nu^{k}h(z,z^{*})$. Those eigenvalues will be of
relevance, which are larger than unity in modulus, and which are
not associated with infinitesimal variable changes.

To solve the eigenproblem numerically we may expand $h(z,z^{*})$
over powers of $z$ and $z^{*}$, and use the polynomial
representation for $g(z)$, which is already known. As we keep a
finite number of terms in the expansions, the eigenproblem reduces
to a finite-dimensional one. Then, the spectrum of eigenvalues may
be calculated by means of standard methods of linear algebra.

As the computations show, the eigenfunction associated with the
senior eigenvalue does not depend on the second argument $z^{*}$.
The coefficients of the power expansion are listed in Table 2;
only even-power terms are nonzero. It is clear that this
eigenfunction represents a perturbation, which retains the map in
the class of complex-analytic functions. The computed eigenvalue
is
\begin{equation}\label{delta1}
\delta_{1}=4.60022558-8.98122473i,
\end{equation}
in good coincidence with the data of Refs.~\cite{Golberg,
Cvitanovic2, Cvitanovic1}.

It appears, however, that one more relevant eigenfunction exists
with eigenvalue larger than unity in modulus, and the polynomial
expansion contains powers of both arguments $z$ and $z^{*}$ (Table
3). This solution is responsible for the non-analytic perturbation
to the fixed point of the RG equation. As we shall see, the
eigenvalue may be expressed explicitly via the scaling constant
$\alpha$.

We know that the fixed-point function $g(z)$ is even. Hence, it
follows from (\ref{iterhnu}) that
\begin{equation}\label{minushnu}
\begin{array}{c}
\nu
h(-z,-z^{*})=\alpha[g'(g(g(z/\alpha)))g'(g(z/\alpha))h(-z/\alpha,-(z/\alpha)^{*})+
\\  +g'(g(g(z/\alpha)))h(-g(z/\alpha),-(g(z/\alpha))^{*})+
h(-g(g(z/\alpha)),-(g(g(z/\alpha)))^{*})].
\end{array}
\end{equation}
We may represent the eigenfunction as
\begin{equation}\label{sanda0}
h(z,z^{*})=h^{s}(z,z^{*})+h^{a}(z,z^{*}),
\end{equation}
where
\begin{equation}\label{sanda1}
h^{s}(z,z^{*})=\frac{h(z,z^{*})+h(-z,-z^{*})}{2},
\end{equation}
is a symmetric part, and
\begin{equation}\label{sanda2}
h^{a}(z,z^{*})=\frac{h(z,z^{*})-h(-z,-z^{*})}{2}.
\end{equation}
is an antisymmetric component. Then, subtracting (\ref{minushnu})
from (\ref{iterhnu}) we have
\begin{equation}\label{itha}
\nu h^{a}(z,z^{*})=\alpha
g'(g(g(z/\alpha)))g'(g(z/\alpha))h^{a}(z/\alpha,(z/\alpha)^{*}).
\end{equation}
Now let us set
\begin{equation}\label{ha}
 h^{a}(z,z^{*})=\frac{g'(z)}{z}\Phi(z,z^{*}).
\end{equation}
We have $g'(z) \propto z$, so the ratio $g'(z)/z$ does not possess
a singularity at the origin. Using the relation
\begin{equation}\label{vg}
g'(g(g(z/\alpha)))g'(g(z/\alpha))g'(z/\alpha)=1,
\end{equation}
which follows from (\ref{RG1}), we obtain a simple equation for
$\Phi(z,z^{*})$:
\begin{equation}\label{nphi}
\nu \Phi (z,z^{*})=\alpha^{2}\Phi ((z/\alpha),(z/\alpha)^{*}).
\end{equation}
Obviously, any product $z^{M}(z^{*})^{N}$ with integer $M \geq 0,
N>0$, $M+N$ odd, represents an eigenfunction, and the associated
eigenvalue is $\alpha^{2-M}(\alpha^{*})^{-N}$. Observe that one of
these numbers is of modulus larger than 1. Indeed, setting $M=0$
and $N=1$, i.e. $\Phi=z^{*}$, we have
\begin{equation}\label{handnu1}
h^{a}(z,z^{*})=\frac{g'(z)}{z}z^{*},
\end{equation}
and
\begin{equation}\label{handnu2}
\nu
=\frac{\alpha^{2}}{\alpha^{*}}=\delta_{2}=2.58728651+1.80679396i.
\end{equation}
The antisymmetric part of the eigenfunction, as well as the
eigenvalue $\delta_{2}$, coincide with those obtained numerically
(Table 3).

All eigenvalues beside $\delta_{1}$ and $\delta_{2}$ are
irrelevant. Some of them are associated with infinitesimal
variable changes (cf.~\cite{Feigenbaum2}). They may be expressed
explicitly via the fixed point function $g(z)$, see Table 4.
Numerical calculations show that all other eigenvalues are of
moduli less than 1.

The condition for realization of the critical behaviour in some
map is a possibility of tuning its parameters to make complex
coefficients at two relevant eigenvectors associated with
$\delta_{1}$ and $\delta_{2}$ equal zero. It yields two complex,
or four real, equations on the parameters. Generically, the
solution will exist if the number of unknowns is not less than the
number of equations. So, we conclude that the phenomenon may occur
as a generic in a space of four real parameters. In other words,
it is of codimension four.

In our line of reasoning we have started from the analytic map.
Nevertheless, analyticity is not a necessary condition for the GSK
critical behaviour. To clarify this remark, let us take the
analytic map $g(z)$ (the fixed point function of the RG equation)
and introduce a non-analytic perturbation, a small term
represented by a smooth function of two arguments, $z$ and $z^*$.
It may be chosen in such a special way that there will be no
contribution from the growing non-analytic eigenmode associated
with the eigenvalue $\delta_2$.(See Eq.(\ref{nphi}) and the
following discussion.) Then, under subsequent iterations of the RG
transformation the perturbation evolving in accordance with
Eq.(\ref{iterh}) will decay. It means that the evolution operator
will tend to the same fixed point function as in the analytic map.

Let us consider a concrete example, the map studied by
Gunaratne~\cite{Gunaratne}
\begin{equation}\label{gunar}
\begin{array}{c}
x'=a+x^{2}-y^{2}, \\ y'=b+(2+\epsilon)xy.
\end{array}
\end{equation}
By means of variable changes $z=x+yi$, $\lambda=a+ib$ and
$\varepsilon = \epsilon/4$ this map can be rewritten as $z' =
\lambda + z^{2} + \varepsilon (z^{2} - (z^{*})^{2})$. Observe that
the non-analytic term  $(z^{*})^{2}$ is quadratic, not linear,
hence, it can not give rise to the eigenvector with eigenvalue
$\delta_2$. As to the contribution into the growing analytical
eigenmode (eigenvalue $\delta_1$), it may be compensated by an
appropriate shift of $\lambda=a+ib$. Thus, the GSK point must
exist; this assertion is in accordance with the conclusion of
Ref.\cite{Gunaratne}. As we have found, for $\epsilon=0.2$ the
critical point is located at $a\cong -0.0355$, $b\cong 0.7477$.


\section{Model map and local scaling coordinates at the GSK critical point}
Now we will construct a model map to study a vicinity of the GSK
critical point in the extended parameter space.

If we consider a shift of $\lambda$ from the critical value in the
complex quadratic map (\ref{logistic}), then only a perturbation
given by the eigenfunction $h_{1}(z)$ can arise. As we wish the
perturbation $h_{2}(z,z^{*})$ to appear, it is necessary to add an
appropriate non-analytic term. This eigenfunction behaves at small
$|z|$ as $h_{2} \propto z^{*}$, hence, we rich this goal adding
the term proportional to $z^{*}$. Thus, we come to the model map
\begin{equation}\label{model}
z_{n+1}=\lambda-z_{n}^{2}+\varepsilon z_{n}^{*},
\end{equation}
which may be regarded as unfolding of the map (\ref{logistic}).
The extended parameter space is the two-dimensional complex space
$\mathrm{C}^{2}: (\lambda,\epsilon)$. Up to a trivial variable
change $(z \rightarrow -z)$ this is the same map that the authors
of Refs.~\cite{Peckham1,Peckham2} have chosen for their extensive
bifurcation analysis. It is equivalent to a four-parameter
two-dimensional real map
\begin{equation}\label{model2d}
x_{n+1}=a-x_{n}^{2}+y_{n}^{2}+ux_{n}+vy_{n}, \qquad
y_{n+1}=b-2x_{n}y_{n}+vx_{n}-uy_{n},
\end{equation}
where we set $z=x+iy$, $\lambda=a+ib$, $\varepsilon=u+iv$.

As we have two relevant eigenfunctions $h_{1}(z)$ and
$h_{2}(z,z^{*})$, the sequence of the evolution operators
generated by iterations of the RG transformation will behave as
\begin{equation}\label{evolution}
f_{k}(z,z^{*})=g(z)+C_{1}(\lambda,\varepsilon)\delta_{1}^{k}h^{(1)}(z)
+C_{2}(\lambda,\varepsilon)\delta_{2}^{k}h^{(2)}(z,z^{*});
\end{equation}
this is true in linear approximation in respect to the deflection
from the fixed point solution $g(z)$. Here $C_{1}$ and $C_{2}$ are
some complex coefficients. They vanish at the GSK critical point
$(\lambda=\lambda_{c},\varepsilon=0)$, but become nonzero as we
depart from it in the parameter space.

It would be convenient to use $C_{1}$ and $C_{2}$ as local
coordinates in a neighborhood of the critical point. In these
coordinates the scaling properties of the parameter space are very
evident. Indeed, let us change $C_{1}$ to $C_{1}/\delta_{1}$ and
$C_{2}$ to $C_{2}/\delta_{2}$. Then, according to
(\ref{evolution}) the evolution operator $f_{k+1}$ will accept the
same form as the operator $f_{k}$ at the old point. It means that
the similar dynamical regimes will occur at the new point, but
with tripled characteristic time scale.

Unfortunately, we do not know explicit expressions of the
coefficients $C_{1}$ and $C_{2}$ via the parameters of the model
map (\ref{model}). Hence, the coordinate system suitable for
demonstration of scaling ({\it the scaling coordinates}) must be
found numerically with a sufficient precision.

Let us consider a manifold $M$ in the parameter space determined
by a condition $C_{1}(\lambda,\varepsilon)=0$, or
$C_{1}(a,b;u,v)=0$, where we set $\lambda=a+ib$ and
$\varepsilon=u+iv$. We can rewrite this equation as
\begin{equation}\label{ab}
a=a_{c}+F(u,v), \qquad b=b_{c}+G(u,v),
\end{equation}
where $a_{c}$ and $b_{c}$ correspond to the GSK critical point.

In fact, we may cut the Taylor expansions for $F(u,v)$ and
$G(u,v)$ keeping the terms of the first and second order. The
reason is the following. As we stay at the manifold $M$, the
renormalization of the complex coordinate $\varepsilon=u+iv$ by
factor $\delta_{2}$ has to ensure realization of similar dynamical
regimes. Let us suppose that we miss some term of power $m$ in the
expansion of $F$ or $G$. When we perform the scale change
$\varepsilon \rightarrow \varepsilon/\delta_{2}^k$, this term will
be of order $\delta_{2}^{-mk}$. So, the error in the amplitude of
the senior eigenvector will behave as
$\delta_{1}^{k}/\delta_{2}^{mk}$. In the case of our concrete type
of criticality (GSK) it is dangerous only for $m \leq 2$,
otherwise the error asymptotically vanishes as $k \rightarrow
\infty$. Indeed, in accordance with (\ref{delta1}) and
(\ref{handnu2}) we have $|\delta_{2}|<|\delta_{1}|$ and
$|\delta_{2}|^{2}<|\delta_{1}|$, but
$|\delta_{2}|^{3}>|\delta_{1}|$. So, only terms of the first and
second order in the Taylor expansion must be evaluated
accurately,\footnote{See other examples of nonlinear parameter
change for observation of the multi-parameter scaling in
Refs.~\cite{Kuznetsov1,Kuznetsov2,Kuznetsov3}.} and we may write
\begin{equation}\label{manif}
\begin{array}{c}
\quad F(u,v)=Au+Bv+Pu^{2}+Quv+Rv^{2},\\ \quad
G(u,v)=Cu+Dv+Su^{2}+Tuv+Uv^{2}.
\end{array}
\end{equation}
Here the capital letters $A$, $B$,...$U$ designate some constants,
which may be found numerically.

As mentioned, precisely at the GSK critical point the map has an
infinite sequence of unstable periodic orbits. Let we come out
this point and look at some cycle of period $N=3^{k}$. In terms of
real perturbation vector $(\tilde{x},\tilde{y})=($Re$\hspace{2pt}
\tilde{z},~$Im$\hspace{2pt} \tilde{z})$ the evolution over one
period is determined by the Jacobi matrix
\begin{equation} \label{jacobi}
\mathbf{J} = \left(
\begin{array}{cc}
 a_{11} & a_{12} \\
 a_{21} & a_{22}
\end{array}
\right)
 = \left(
\begin{array}{cc}
  \partial x_{N} / \partial x_{0} & \partial x_{N} / \partial y_{0} \\
  \partial y_{N} / \partial x_{0} & \partial y_{N} / \partial y_{0}
\end{array}
\right).
\end{equation}
While the map is complex analytic, the Cauchy -- Riemann equations
hold, and $a_{11}=a_{22}$, $a_{12}=-a_{21}$. Otherwise, in
presence of the term $\varepsilon z^{*}$, these equalities are
violated.

Let us consider trace
\begin{equation}\label{trace}
S=a_{11}+a_{22}
\end{equation}
and determinant
\begin{equation}\label{determ}
J=a_{11}a_{22}-a_{12}a_{21}
\end{equation}
of the matrix $\mathbf{J}$ at the cycle. Surely, these values
depend on the parameters of the map:
$S=S(\lambda,\varepsilon)=S(a,b,u,v)$ and
$J=J(\lambda,\varepsilon)=J(a,b,u,v)$. Precisely at the critical
point $(a=a_{c},h=b_{c},u=0,v=0)$ they tend, respectively, to
$2\hspace{1pt}$Re$\hspace{2pt}\mu_{c}$ and $|\mu_{c}^{2}|$, as $k$
grows.

Let us suppose that we know the first and the second derivatives
of the trace and determinant in respect to the parameters
$u=$Re$\hspace{2pt} \varepsilon$ and $v=$Im$\hspace{2pt}
\varepsilon$ for two sufficiently large periods, $3^{k+1}$ and
$3^{k}$. We require the following scaling property to hold. If one
takes (i) some infinitesimal value of
$\varepsilon=\varepsilon_{0}=u+iv$ and (ii) the rescaled value
$\varepsilon=\delta_{2}\varepsilon_{0}=(\delta'_{2}u-\delta''_{2}v)
+i(\delta'_{2}v+\delta''_{2}u)$, where $\delta'_2=$Re$\hspace{2pt}
\delta_2$, $\delta''_2=$Im$\hspace{2pt} \delta_2$, then traces and
determinants for both cycles must coincide at arbitrarily chosen
$u$ and $v$:
\begin{equation}\label{s12}
\begin{array}{c}
S^{k+1}(u,v)=S^{k}(\delta'_{2}u-\delta''_{2}v,\delta'_{2}v+\delta''_{2}u),
\\
J^{k+1}(u,v)=J^{k}(\delta'_{2}u-\delta''_{2}v,\delta'_{2}v+\delta''_{2}u).
\end{array}
\end{equation}
Now we substitute the Taylor expansion $S(u,v)=S_{0}+S_{u}u+S_{v}v
+\frac{1}{2}S_{uu}u^{2}+S_{uv}uv+\frac{1}{2}S_{vv}v^{2}$, and
analogous representation for $J(u,v)$, and equalize separately
coefficients at the terms of the first and of the second power in
$u$ and $v$. It yields ten real equations with ten unknowns
$A,B,...U$. We have found numerically the derivatives
$(S_{u},S_{v},S_{uu},..., J_{u},J_{v},J_{uu},...)$ for
sufficiently large periods and solved the mentioned equations. The
resulting values of the constants were found to be asymptotically
stable, estimated as
\begin{equation}\label{koefscal1}
\begin{array}{c}
A=-0.232750391, B=0.02699484,C=-B, D=A, \\
 P=-0.2173, Q=-0.070, R=-0.3184, \\
 S=0.1533, T=0.1010, U=0.0830.
\end{array}
\end{equation}
(In our computations the maximal periods we could handle were
about $3^8$ or $3^9$. However, the convergence is rather fast, and
already at $3^4\div 3^7$ we obtain sufficiently accurate data.)

Then, the the model map is represented as
\begin{equation}\label{scal1}
\begin{array}{c}
x'=a-x^{2}+y^{2}+ux+vy+Au+Bv+Pu^{2}+Quv+Rv^{2}, \\
y'=b-2xy+vx-uy+Cu+Dv+Su^{2}+Tuv+Uv^{2},
\end{array}
\end{equation}
and the coordinate system $(a,b,u,v)$ is appropriate for the
demonstration of scaling. Indeed, a shift of $u$ and $v$ will
contribute exceptionally into the second eigenvector, and a shift
of $a$ and $b$ --- only into the first eigenvector.
\section{Scaling properties of the extended parameter space in a
neighborhood of the critical point}

The number of real parameters in the model map (\ref{scal1}) is
four, and complete bifurcation analysis of the parameter space
$(a,b,u,v)$ would be complicated and tedious (see, nevertheless,
Refs.~\cite{Peckham1,Peckham2}). Here we only want to present
computer illustrations for scaling properties following from the
results of Sec.2. We will consider some two-dimensional
cross-sections of the parameter space and look at the "topography
charts" for dynamical regimes on these surfaces.

For graphical presentation we will use the technique of Lyapunov
charts (see~\cite{Rossler,Marcus,Bastos} for previous applications
of this method). As we have chosen the parameter-space
cross-section to be studied, we compute the senior Lyapunov
exponent for the model map at each pixel of the two-dimensional
plot, and mark this pixel by an appropriate grey tone. We code the
Lyapunov exponent from large negative values to zero by tones from
dark to light grey (it corresponds to periodic regimes). White
color represents zero Lyapunov exponent (e.g. quasiperiodicity or
states at the onset of chaos), and black --- positive values
(chaos). Such a convention ensures a clear vision of the border
between regular and chaotic dynamics. In some domains of the
parameter space the model map manifests divergence; these areas
are colored uniformly by one special grey tone.

In our computations the trajectories are launched from the origin
$z=0$, but the calculations for evaluation of the Lyapunov
exponent start after some sufficient number of iterations to
exclude the transients.

To begin, let us illustrate the kind of scaling that follows from
results of Refs.~\cite{Golberg,Cvitanovic2,Cvitanovic1}. In the
top part of Fig.2(a) we show a cross-section of the parameter
space by surface $u=0$, $v=0$, at which the equation (\ref{scal1})
reduces to a conventional complex analytic quadratic map
(\ref{logistic}). What is observed is just a fragment of the
Mandelbrot cactus. The GSK critical point is located exactly at
the center of the diagram. We select a small box containing the
critical point, enlarge and rotate it in accordance with
multiplication by $\delta_{1}\cong 4.6002-8.9812i$, and then
reproduce on the bottom plot. In a course of this procedure we
redefine the legend for the Lyapunov exponent: as the
characteristic time scale triplicates, we decrease by $3$ times
all the values separating the intervals for coding by definite
grey tones. Observe excellent visual similarity of both pictures.

Next, let us take another cross-section of the parameter space of
the map (\ref{scal1}), by the surface $(a=0,b=0)$. It means that
in the parameter space of the original map (\ref{model}) we are at
the manifold $M$: $C_{1}(\lambda,\varepsilon)=0$. The Lyapunov
chart is shown in Fig.2(b); the critical point is placed precisely
in the middle of the plot. Observe that visually the picture has
nothing common with the familiar Mandelbrot set. However, the
self-similarity does hold, although it is governed now by the new
scaling constant (\ref{handnu2}). To demonstrate it, we take a box
around the critical point. After magnification and rotation
corresponding to multiplication by $\delta_{2}\cong
2.5872+1.8067i$, and with redefinition of the grey-scale coding
rule for the Lyapunov exponent, we obtain the chart shown on the
bottom plot. It looks remarkably similar to the top diagram.

To give clearer impression of the parameter space arrangement on
the manifold $M$ we present in Fig.3 a schema explaining some
details of the bifurcation structure. The largest area in the
middle corresponds to a stable period-$1$ state. Here the map
possesses the fixed point with two complex conjugate multipliers
of modulus less than unity. From the bottom this domain is bounded
by a curve, at which the Neimark bifurcation occurs. Here two
complex multipliers cross the unit circle. Along the border
argument of complex multiplier is varied. For irrational
arguments, measured in $2\pi$ units, an attractor, which appears
as a result of the bifurcation, will be a torus. For rational
values it is a resonance cycle (at the sharp ends of the Arnold
tongues). At the upper border of the stability domain
period-doubling bifurcation takes place. It may be followed either
by secondary period doubling (top left part of the border for the
period-$2$ domain), or by Neimark bifurcation (top right part of
the border). Note overlap of the period-$1$ and period-$2$ areas,
which indicates presence of multistability (coexistence of two
attractors with distinct basins). At this region the borders of
the stability domains appear to be associated with the hard
bifurcations (jumps). At the left bottom part of this structure
another similarly arranged smaller formation is attached. Here the
period-3 regime occurs in the middle part of the stability domain.
According to results of the RG analysis and scaling arguments, the
sequence of the smaller domains, each of which is attached to its
precursor, must be infinite. Asymptotically they become
self-similar, accepting a universal form. (In fact, already on the
very crude resolution scale of Fig.3 the domains of periods one,
three, and nine look remarkably similar.)

The present illustrations support our basic statement that the
second eigenvalue $\delta_{2}$ is of relevance for the
non-analytic modification of the map. The scaling properties of
the extended parameter space are determined by two constants, the
earlier known $\delta_{1}$, and the novel $\delta_{2}$.

\section{Conclusion}
In this article we discuss scaling behaviour of complex maps at
the accumulation point of period-tripling in the extended
parameter space.

We consider the fixed point function of the RG equation of
Golberg, Sinai, and Khanin, and analyze a class of perturbations,
which are supposed to be small and smooth, but not necessarily
obey the Cauchy-Riemann conditions. There is no need in
modification of the fixed point equation, but we revise the
linearized RG equation, which describes behaviour in a vicinity of
the analytic fixed point. Then, we study the eigenvalue spectrum
and detect two relevant eigenvectors, one of which is analytic,
and another one is non-analytic and violates the Cauchy -- Riemann
conditions.

To observe the period-tripling universal scaling behaviour in a
non-analytic map one needs to make {\it two} complex coefficients
at the relevant eigenvectors equal zero by means of appropriate
selection of parameters. Typically, it requires controlling not
less than four real parameters. It may be thought that the
possibility of observation for such type of behaviour in physical
systems is rather problematic. Nevertheless, we can not exclude
that the period-tripling criticality may be found in some systems,
which possess special symmetries. Apparently, this is an
interesting direction of search for physical applications of the
complex analytic dynamics.

\section{Acknowledgement}

This work is supported, in part, by RFBR, grant 00-02-17509. S.K.
acknowledges support from Max Planck Society and Max Planck
Institute of Physics of Complex Systems in Dresden, where a part
of this work has been done during workshop and seminar
"Statistical Mechanics of Space-Time Chaos" (July 2000). O.I.
acknowledges support from RFBR, grant 01-02-06385. S.K. and O.I.
acknowledge support via Research and Educational Center of
Nonlinear Dynamics and Biophysics of Saratov State University from
CRDF, award No REC-006.


\newpage

\begin{table}
\begin{center}
\caption{The coefficients of polynomial expansion for the
fixed-point function $g(z)$}
\begin{tabular}{|c|c|}
\hline

1 & $1.0 + 0.0i$ \\ \hline

$z^{2}$ & $0.054665304 - 0.749020944i$ \\ \hline

$z^{4}$ & $- 0.024397241 - 0.052466461i$ \\ \hline

$z^{6}$ & $- 0.002529112 - 0.001197430i$ \\ \hline

$z^{8}$ & $- 0.000088081 + 0.000137556i$ \\ \hline

$z^{10}$ & $0.000000729 + 0.000018289i$ \\ \hline

$z^{12}$ & $0.000000541 + 0.000001194i$ \\ \hline

$z^{14}$ & $0.000000074 + 0.000000048i$ \\ \hline

\hline
\end{tabular}
\end{center}
\end{table}

\newpage
\begin{table}
\begin{center}
\caption{The coefficients of polynomial expansion for the senior
eigenfunction $h^{(1)}(z)$}

\begin{tabular}{|c|c|}
\hline

1 & $1.0 + 0.0i$ \\ \hline

 $z^{2}$ & $0.181223377 - 0.141361034i$ \\ \hline

$z^{4}$ & $0.002303322 - 0.015962515i$ \\ \hline

$z^{6}$ & $- 0.001131785 - 0.000491487i$ \\ \hline

$z^{8}$ & $- 0.000151737 + 0.000053131i$ \\ \hline

$z^{10}$ & $- 0.000010514 + 0.000009686i$ \\ \hline

$z^{12}$ & $- 0.000000310 + 0.000001085i$ \\ \hline

$z^{14}$ & $0.000000033 + 0.000000100i$ \\ \hline
\end{tabular}
\end{center}
\end{table}

\newpage
\begin{table}
\begin{center}
\caption{The coefficients of polynomial expansion for the
eigenfunction $h^{(2)}(z,z^{*})$ associated with the eigenvalue
$\delta_{2}=2.58728651+1.80679396i$}
\begin{tabular}{|c|c|c|c|c|c|c|}
\hline

  & $1$ & $z^{*}$ & $(z^{*})^{2}$ & $(z^{*})^{4}$ & $(z^{*})^{6}$ &
 $(z^{*})^{8}$ \\
\hline

 $1$ & $1.0+0.0i$ & $-3.029846$ & $0.398114$ &
$-0.022670$ & $-0.002329$ & $-0.000073$
\\

 &  & $+6.351031i$ & $+0.008514i$ &
$-0.073276i$ & $-0.003512i$ & $-0.0002278i$\\ \hline

 $z^{2}$ & $1.094504$ & $0.068249$ & $0.086111$ &
$-0.001995$ & $-0.000053$ & $0.0000004$ \\

 & $-1.043125i$ & $+1.082134i$ & $+0.039504i$ & $+0.001278i$ &
$+0.000130i$ & $+0.000006i$ \\ \hline

$z^{4}$ & $-0.048082$ & $0.053978$ & $0.001900$ & $-0.000124$ &
$0.0000003$ &  \\

 & $-0.141290i$ & $+0.057211i$ & $-0.001493i$ & $+0.000212i$ &
$+0.000009i$ &  \\ \hline

 $z^{6}$ & $-0.011263$ & $0.004950$ &
$-0.000450$ & $-0.000005$ &   &  \\

 & $-0.005046i$ & $-0.003601i$ & $-0.000460i$ & $+0.000022i$ &
&  \\ \hline

$z^{8}$ & $-0.000616$ & $0.000280$ & $-0.000058$ & $0.0000002$ & &
\\

 & $+0.000710i$ & $-0.000811i$ & $-0.000028i$ & $+0.000002i$ &
&    \\ \hline

$z^{10}$ & $-0.000002$ & $-0.000004$ & $-0.000004$ & &    &
\\

 & $+0.000109i$ & $-0.000074i$ & $+0.0000002i$ & & &   \\

\hline
\end{tabular}
\end{center}
\end{table}

\begin{table}
\begin{center}
\caption {Some eigenfunctions of the eigenproblem (\ref{iterhnu})
associated with the infinitesimal variable changes}
\begin{tabular}{|c|c|c|}
\hline

Variable change, & Eigenfunction & Eigenvalue $\nu$
\\
$\Delta \ll 1$ &  $h_{2}(z,z^{*})$ &
\\
\hline

Shift: & $g'(z)-1$ & $\nu=\alpha =-2.0969+2.3583i$
\\
$z \rightarrow z + \Delta$ &  &
\\
\hline

Scale change  & $z g'(z)-g(z)$ & $\nu=1$
\\
$z \rightarrow z(1 + \Delta)$ &  &
\\
\hline

Non-analytic change & $z^{*} g'(z)-(g(z))^{*}$ & $\nu=\alpha
/\alpha^{*}=-0.1169-0.9931i$
\\
$z \rightarrow z + z^{*} \cdot \Delta$ & & $|\nu|=1$
\\
\hline

\end{tabular}

\end{center}
\end{table}

\newpage
\begin{figure}
\centerline{\includegraphics[scale=1.6]{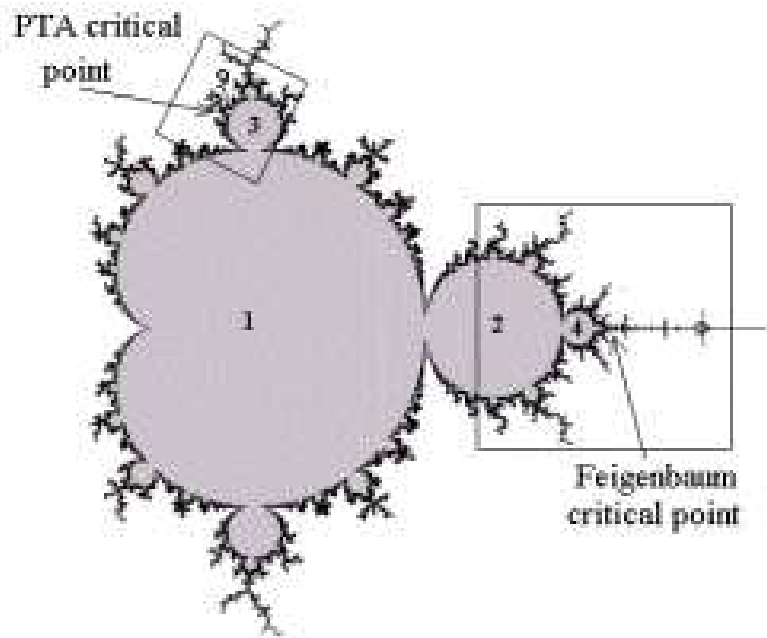}}

\caption{Mandelbrot set for complex analytic quadratic map
(\protect\ref{logistic}). A unification of grey colored areas is
the Mandelbrot cactus. Periods associated with the leaves of the
cactus are marked by respective numbers. Two selected boxes show
domains of the Feigenbaum period-doubling scaling and of the
period-tripling scaling.}
\end{figure}

\newpage
\begin{figure}
\centerline{\includegraphics[scale=1.6]{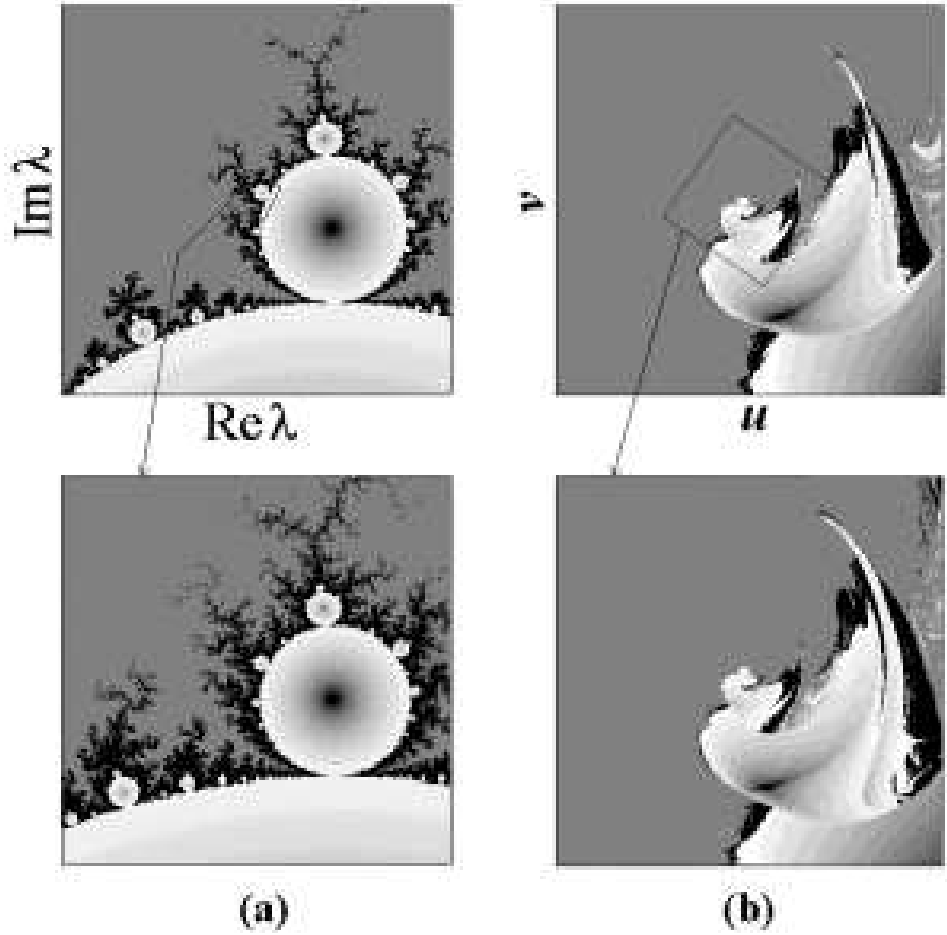}}

\caption{Cross-section of the parameter space by a surface $u=0$,
$v=0$, at which the map (\protect\ref{scal1}) is complex analytic
(a), and by the manifold $M$, at which the contribution of the
senior eigenvector into a perturbation of the RG fixed point is
excluded (b). Grey tones from dark to light code values of the
Lyapunov exponent from large negative to zero. White color
corresponds to zero, and black -- to positive Lyapunov exponent.
Divergence is shown by uniform coloring with one special grey
tone. The GSK critical point is located exactly at the center of
the diagrams. A small box containing the critical point, is
magnified and rotated in accordance with multiplication by
$\delta_{1} \simeq 4.6002-8.9812i$ on the panel (a) and by
$\delta_{2} \simeq 2.5873+1.8068i$ on the panel (b). The coding
rule on the bottom plots is redefined to reveal the
self-similarity: all the values separating intervals for coding by
definite grey tones are decreased by 3 times.}
\end{figure}

\newpage
\begin{figure}
\centerline{\includegraphics[scale=1.6]{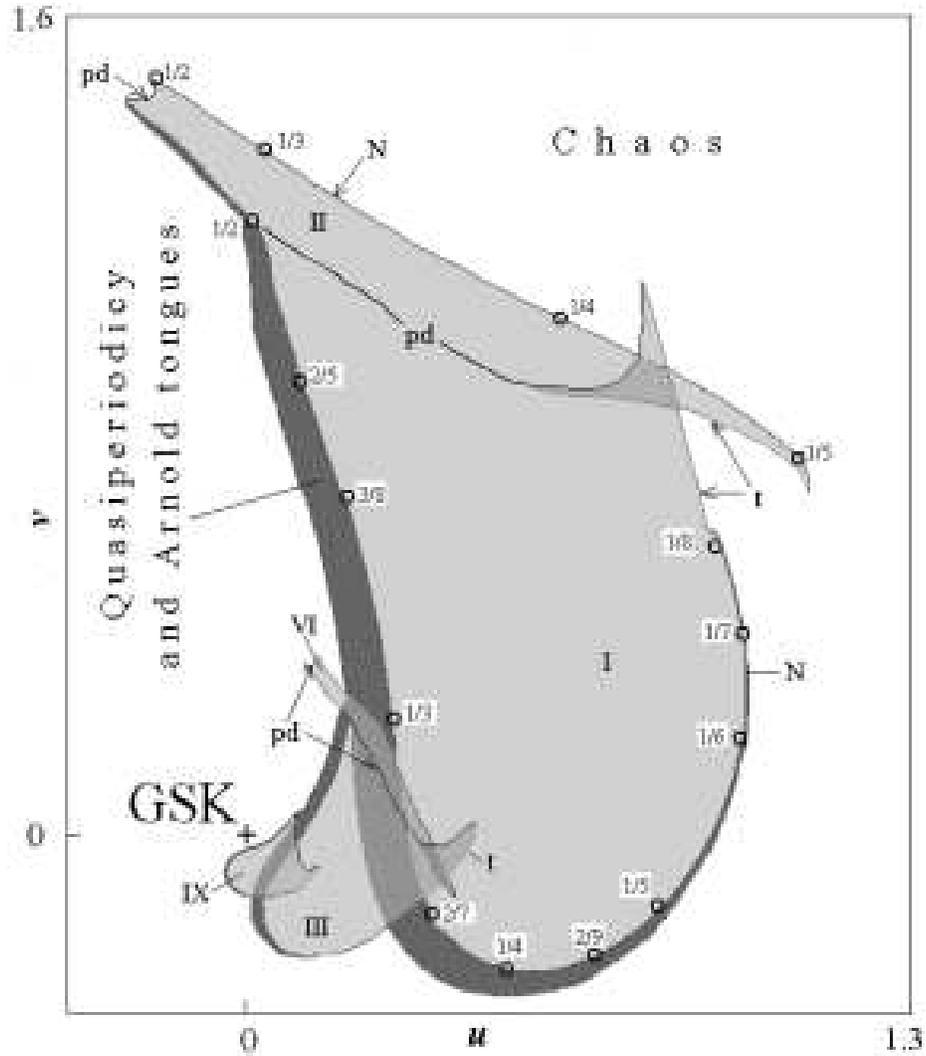}}

\caption{A schema of the parameter space arrangement in the
cross-section of the extended parameter space by the manifold $M$,
at which the contribution of the senior eigenvector into a
perturbation of the RG fixed point is excluded. Bifurcation lines
are marked as $pd$ (period-doubling), $t$ (tangent), $N$
(Neimark). Light grey areas correspond to periodic regimes of
periods marked by the Roman figures. Dark grey designates the area
of quasi-periodicity and Arnold tongues (not shown). On the
Neimark bifurcation curve some points of rational arguments of the
multiplier are marked, these are the sharp ends of the respective
Arnold tongues. The GSK critical point is marked by a cross.}
\end{figure}

\end{document}